# Dissociated Neuronal Cultures as Model Systems for Self-Organized Prediction

Amit Yaron[1], Zhuo Zhang[2], Dai Akita[2], Tomoyo Isoguchi Shiramatsu[2], Zenas Chao[1], Hirokazu Takahashi[1,2*]

[1] International Research Center for Neurointelligence (WPI-IRCN), The University of Tokyo Institutes for Advanced Study (UTIAS), The University of Tokyo, Tokyo, Japan

[2] Department of Mechano-Informatics, Graduate School of Information Science and Technology, The University of Tokyo, Tokyo, Japan

**\* Correspondence:**
Corresponding Author
Hirokazu Takahashi (takahashi@i.u-tokyo.ac.jp)

**Abstract**

Dissociated neuronal cultures provide a simplified yet effective model system for investigating self-organized prediction and information processing in neural networks. This review consolidates current research demonstrating that these in vitro networks display fundamental computational capabilities, including predictive coding, adaptive learning, goal-directed behavior, and deviance detection. We examine how these cultures develop critical dynamics optimized for information processing, detail the mechanisms underlying learning and memory formation, and explore the relevance of the free energy principle within these systems. Building on these insights, we discuss how findings from dissociated neuronal cultures inform the design of neuromorphic and reservoir computing architectures, with the potential to enhance energy efficiency and adaptive functionality in artificial intelligence. The reduced complexity of neuronal cultures allows for precise manipulation and systematic investigation, bridging theoretical frameworks with practical implementations in bio-inspired computing. Finally, we highlight promising future directions, emphasizing advancements in three-dimensional culture techniques, multi-compartment models, and brain organoids that deepen our understanding of hierarchical and predictive processes in both biological and artificial systems. This review aims to provide a comprehensive overview of how dissociated neuronal cultures contribute to neuroscience and artificial intelligence, ultimately paving the way for biologically inspired computing solutions.

**1 Introduction**

The brain's remarkable ability to process information, learn from experience, and adapt to changing environments emerges from the dynamic interactions of billions of neurons. Understanding how these capabilities arise from neural network organization represents a fundamental challenge in neuroscience (Friston et al., 2006; Friston, 2010; Bastos et al., 2012; Keller and Mrsic-Flogel, 2018). Dissociated neuronal cultures—simplified systems where neurons are isolated from their native environment and allowed to self-organize—provide a powerful experimental platform for investigating these processes. These cultures retain core capabilities for network formation, information processing, and adaptation while offering unprecedented access for manipulation and observation (Maeda et al., 1995; Kamioka et al., 1996; Potter and DeMarse, 2001; Marom and Shahaf, 2002).

The study of neuronal cultures has evolved dramatically since Ross Granville Harrison first demonstrated nerve fiber growth in vitro in 1910 (Harrison, 1910). Harrison's pioneering work established the foundation for modern neurobiology by enabling direct observation of neural development. A transformative advance came with the introduction of microelectrode array (MEA) technology (Figure 1A) (Thomas et al., 1972; Gross et al., 1977; Pine, 1980). MEAs revolutionized the field by enabling long-term, non-invasive recording from multiple neurons simultaneously, providing unprecedented insight into network dynamics and development (Pine, 2006; Bakkum et al., 2013; Müller et al., 2015; Obien et al., 2015). Early MEA platforms allowed researchers to monitor network formation in dissociated cultures, revealing spontaneous activity and plasticity (Figures 1B and 1C). High-density CMOS microelectrode arrays now enable recording from thousands of neurons with unprecedented spatial and temporal resolution (Berdondini et al., 2005; Frey et al., 2007; Ballini et al., 2014; Müller et al., 2015). These systems (Figure 1D) facilitate detailed investigations of both localized interactions and long-range network dynamics. They provide subcellular resolution, as illustrated by the precise alignment of neurons with individual electrodes (Figure 1E) and allow spatial mapping of extracellular spikes overlaid on neuronal morphology to track activity sources and connectivity (Figure 1F). Research using these systems has revealed several fundamental properties of neural network organization and function.

Research using these systems has revealed several fundamental properties of neural network organization and function. As cultures develop, they demonstrate a remarkable capacity for self-organization, evolving from random collections of cells into functional networks that exhibit critical dynamics optimized for information (Beggs and Plenz, 2003; Levina et al., 2007; Millman et al., 2010; Friedman et al., 2012; Yada et al., 2017; Kossio et al., 2018). These networks show robust capabilities for learning and memory formation, as demonstrated through studies of synaptic plasticity and adaptive responses to electrical stimulation (Jimbo et al., 1998, 1999; Shahaf and Marom, 2001; Le Feber et al., 2010, 2014, 2015; Dranias et al., 2013; Dias et al., 2021)

Neuronal cultures have proven effective for studying goal-directed behavior in closed-loop systems. Potter et al. (1997) introduced the 'Animat in a Petri Dish' concept, establishing a paradigm where network activity controlled a simulated animal ('animat') while receiving sensory feedback through electrical stimulation. This foundational work led to numerous studies demonstrating that cultured networks can adapt to control external devices (DeMarse et al., 2001; Potter et al., 2003; Bakkum et al., 2008b; Chao et al., 2008a; Tessadori et al., 2012; Masumori et al., 2020; Yada et al., 2021; Kagan et al., 2022), advancing our understanding of neural adaptation and control while suggesting new approaches for brain-machine interfaces and neuroprosthetics.

The computational capabilities of neuronal cultures extend to more sophisticated information processing tasks. These networks exhibit predictive coding and deviance detection, supporting theoretical frameworks such as the free energy principle (Rao and Ballard, 1999; Friston, 2010; Huang and Rao, 2011; Isomura et al., 2015; Isomura and Friston, 2018; Lamberti et al., 2023). Their ability to perform complex computations while maintaining remarkable energy efficiency has important implications for neuromorphic computing and artificial intelligence (Marković et al., 2020; Smirnova et al., 2023). Insights from neuronal cultures can influence the development of new computing architectures, particularly in areas such as reservoir computing and adaptive neural networks (Dockendorf et al., 2009; Kubota et al., 2019, 2021; Tanaka et al., 2019; Subramoney et al., 2021; Cai et al., 2023; Sumi et al., 2023).

The development of three-dimensional culture techniques and brain organoids offers new opportunities to study neural organization in more physiologically relevant contexts (Hogberg et al., 2013; Lancaster et al., 2013; Clevers, 2016; Smirnova and Hartung, 2024). These advances, combined with sophisticated analysis techniques and theoretical frameworks, are providing new insights into how neural networks self-organize for efficient information processing and adaptation.

This review synthesizes current research on dissociated neuronal cultures, examining their contributions to our understanding of neural network organization and function. We begin by exploring network development and the emergence of critical dynamics (Chapter 2), followed by detailed analysis of learning and memory formation in these systems (Chapter 3). We then examine how neuronal cultures exhibit deviance detection and predictive processing (Chapter 4), and their remarkable capacity for goal-directed behavior when coupled with external systems (Chapter 5). The relationship between these empirical findings and theoretical frameworks, particularly the free energy principle, is explored in Chapter 6. Chapter 7 discusses how insights from neuronal cultures inform the development of artificial neural networks and neuromorphic computing systems. Finally, we consider future directions for the field, including advances in three-dimensional culture techniques, brain organoids, and their implications for both neuroscience and artificial intelligence (Chapter 8).

By examining how these simplified neural systems self-organize for prediction and adaptation, we aim to illuminate fundamental principles of neural computation while highlighting their practical applications in bio-inspired computing and neuroprosthetics. This understanding may ultimately guide the development of more efficient and adaptive artificial systems while deepening our knowledge of biological neural network function.

## 2. Network Development and Self-Organized Criticality

The transformation of dissociated neuronal cultures from random collections of neurons into sophisticated, functionally organized systems is a remarkable feat of biological self-organization. This chapter explores the key processes and principles underlying network development in these cultures, followed by an introduction to the concept of Self-Organized Criticality (SOC) and its relevance to understanding network maturation.

Network development in neuronal cultures progresses through several distinct stages, each characterized by increasingly complex patterns of activity. In the earliest stages, neurons exhibit seemingly chaotic, independent firing patterns. Kamioka et al. (1996) observed that this apparent randomness quickly gives way to more organized activity as the culture matures. The transition from independent firing to coordinated activity is heavily dependent on NMDA receptor activation and is influenced by external factors such as calcium concentrations (Segev et al., 2001). As development continues, the network establishes stable, recurring patterns of synchronized activity. Van Pelt et al. (2004) documented the emergence of network bursting as a hallmark of culture maturation. Further refinement of connections leads to more sophisticated firing patterns, including what Wagenaar et al. (2006) termed "superbursts" - periods of intense, coordinated activity that reflect the increasing complexity of network interactions. As shown by Yada et al. (2017), these patterns exhibit state-dependent properties, with different spatiotemporal patterns appearing successively and periodically, suggesting organized fluctuations in neural activity propagation. **Figure 2** illustrates these developmental transitions using data from high-density CMOS microelectrode arrays. **Figure 2A** displays spatial maps of action potential amplitudes recorded at different developmental stages, while **Figure 2B** highlights changes in spike waveforms at selected electrodes over time. **Figure 2C** depicts the progression of spontaneous spiking activity, showcasing the emergence of synchronized bursts. **Figure 2D** visualizes the shift in neuronal avalanche size distributions, from exponential at early stages (4 DIV) to power-law distributions indicative of SOC by 16 DIV. Lastly, **Figure 2E** presents the integration-fragmentation model explaining SOC emergence, highlighting the role of synaptic pruning and balanced excitation-inhibition dynamics in this transition. The structural and functional organization of the network evolves in parallel with these changes in activity patterns. Over time, synaptic connections become more stable, as evidenced by metrics like conditional firing probabilities (Le Feber et al., 2007). Soriano et al. (2008) demonstrated that maturing networks form modular structures with hierarchical organization, which is crucial for supporting coherent and synchronized activity. This organization has been shown to develop into rich-club topology, which supports coordinated dynamics and information processing (Schroeter et al., 2015; Nigam et al., 2016). Baruchi et al. (2008) characterized how mutual synchronization emerges between coupled networks, demonstrating that despite engineering similarity, spontaneous asymmetries emerge in both activity propagation and functional organization. Intriguingly, Orlandi et al. (2013) discovered that intrinsic noise plays a constructive role in network formation. Through a process they termed "noise focusing," weak, randomly distributed synaptic connections are enhanced, facilitating the transition from disorganized activity to coherent network dynamics.

As researchers sought to understand the principles governing these complex developmental dynamics, the concept of Self-Organized Criticality (SOC) emerged as a powerful explanatory framework. Introduced by Per Bak (Bak et al., 1987; Bak, 1996) in the context of physical systems, SOC describes how complex systems naturally evolve toward a critical state characterized by scale-invariant behavior. Further theoretical work has expanded our understanding of SOC in neural systems, highlighting its ubiquity across different scales of brain organization and its functional implications (Muñoz, 2018; Plenz et al., 2021). Networks at criticality exhibit maximized dynamic range, optimally responding to the broadest range of stimulus intensities (Shew et al., 2009). This concept has been widely applied to neural systems, offering insights into network development and function (Chialvo, 2010; Beggs and Timme, 2012; Shew and Plenz, 2013; Bilder and Knudsen, 2014). In the context of neuronal networks, SOC is most notably manifested in the phenomenon of neuronal avalanches - cascades of spontaneous activity that follow power-law size distributions. Beggs and Plenz (2003) were among the first to observe and characterize these avalanches in neuronal cultures, followed by confirmations in various neural systems (Mazzoni et al., 2007; Pasquale et al., 2008; Petermann et al., 2009; Friedman et al., 2012). Experimental evidence for the development of SOC has come from several studies using advanced recording techniques. Yada et al. (2017) used high-density CMOS microelectrode arrays to capture the progression of avalanche dynamics across three distinct phases: an initial exponential distribution, a transitional bimodal distribution, and a final power-law distribution characteristic of a critical state. This observed sequence supports a gradual expansion model of network development, where neural connections are extended incrementally over time. Kayama et al. (2019) revealed the formation of functional clusters within maturing cultures, showing how these clusters exhibit diverse and repeatable patterns of synchronized firing, indicating the development of specialized subnetworks within the larger network structure. These findings complement earlier observations (Baruchi et al., 2008) about the emergence of mutual synchronization in coupled networks, demonstrating how spontaneous asymmetries arise in both activity propagation and functional organization.

The emergence of SOC in neuronal cultures involves multiple mechanisms developing over time. Vreeswijk and Sompolinsky (1996) demonstrated the importance of balanced excitation and inhibition in neural networks, showing its relationship with chaotic dynamics. Abbott and Rohrkemper (2007) proposed a growth-based mechanism where neurons add or remove synapses based on their activity levels. Both short-term and long-term plasticity contribute to the network's evolution toward criticality (Levina et al., 2007; Millman et al., 2010). Vogels et al. (2011) showed how inhibitory plasticity maintains excitation-inhibition balance in memory networks, and Hennequin et al. (2017) synthesized how inhibitory synaptic plasticity acts as a crucial control mechanism for network stability and computation. The interaction between plasticity mechanisms is particularly important: excitatory STDP with an asymmetric time window destabilizes the network toward a bursty state, while inhibitory STDP with a symmetric time window stabilizes the network toward a critical state (Sadeh and Clopath, 2020). Structural changes, such as axonal elongation and synaptic pruning, also shape the network's critical dynamics (Tetzlaff et al., 2010; Kossio et al., 2018). Kuśmierz et al. (2020) demonstrated that networks with power-law distributed synaptic strengths exhibit a continuous transition to chaos. The relationship between criticality and the edge of chaos represents another important regulatory point in neural networks, associated with the balance between excitation and inhibition. SOC, the edge of chaos, and excitation-inhibition balance serve as complementary homeostatic set points in well-tuned networks, each contributing to the optimization of computation and memory formation. Ikeda et al. (2023) have shown how the interplay between environmental noise and spike-timing-dependent plasticity can drive networks toward criticality, emphasizing the importance of optimal noise levels in this process. Theoretical modeling by Kern et al. (2024 (under review)) has emphasized the crucial role of inhibitory circuitry, demonstrating how the density and range of inhibitory synaptic connections significantly influence the development of critical dynamics.

The study of network development through the lens of SOC has provided valuable insights into the fundamental principles governing the maturation of neuronal systems. It offers a framework for understanding how complex, functional network structures emerge from initially disordered collections of neurons, and how these networks maintain a balance between stability and flexibility as they mature. This self-organized development toward criticality, supported by various plasticity mechanisms and carefully regulated by inhibitory circuits, enables neuronal networks to achieve optimal information processing capabilities while maintaining adaptability. The resulting networks exhibit a rich repertoire of dynamics that supports their computational functions while preserving the ability to respond to changing environmental demands.

**3. Adaptive Learning and Memory Formation**

Dissociated neuronal cultures offer a simplified system for studying learning and memory, providing insight into how neural networks adapt in response to external stimuli. This chapter reviews key findings demonstrating that these cultures exhibit learning behaviors and explores the mechanisms that enable memory formation and adaptation in these systems.

Early studies laid the foundation for understanding learning in dissociated cultures. Jimbo et al. (1999) showed that localized tetanic stimulation could induce potentiation and depression in specific pathways, highlighting the network's capacity to modify connections based on stimuli. Shahaf and Marom (2001) demonstrated that networks could be trained to produce specific responses through low-frequency electrical stimulation, without the need for external reward mechanisms, suggesting that learning can emerge from simple, self-organizing principles. Ruaro et al. (2005) further established the computational capabilities of these cultures, showing they could perform pattern recognition tasks through targeted electrical stimulation. Their work demonstrated how biological neurons could be trained to recognize specific spatial patterns, with responses enhanced through long-term potentiation mechanisms.

Later work explored how network dynamics could be controlled and shaped through stimulation. Wagenaar et al. (2005) demonstrated that closed-loop, distributed electrical stimulation could effectively transform burst-dominated activity into dispersed spiking patterns more characteristic of in vivo activity. Le Feber et al. (2010) showed that adaptive electrical stimulation—where stimulation is adjusted based on network feedback—was more effective at inducing long-lasting connectivity changes compared to random stimulation. This highlighted the role of feedback in shaping the learning process.

Memory formation in dissociated cultures was further investigated by Le Feber et al. (2015), who found that repeated stimulation could create multiple parallel memory traces. This indicated that these cultures could handle complex memory storage tasks, with distinct stimuli producing stable patterns of connectivity. Additionally, Bakkum et al. (2008a) demonstrated that even when synaptic transmission was blocked, changes in action potential propagation still occurred, suggesting that non-synaptic mechanisms contribute to network adaptation.

Short-term memory processes were explored by Dranias et al. (2013), who identified two types of STM in these networks: "fading memory," reliant on reverberating neural activity, and "hidden memory," which persists through changes in synaptic strength even after neural activity has ceased. Ju et al. (2015) expanded on these findings, demonstrating that dissociated networks possess an intrinsic capacity for spatiotemporal memory lasting several seconds and can classify complex temporal patterns. Their work highlighted the importance of short-term synaptic plasticity and recurrent connections in enabling these computational capabilities.

Further studies have provided more detail on the molecular and network dynamics underlying memory and learning. Dias et al. (2021) found that memory consolidation in these cultures was influenced by network state, with low cholinergic tone enhancing memory formation. Ikeda et al. (2021) demonstrated the flexibility of dissociated networks, showing that low-frequency stimulation could initially induce depression but later lead to potentiation, revealing the dynamic nature of learning.

These findings demonstrate that dissociated neuronal cultures are capable of both learning and memory formation through various mechanisms, including synaptic plasticity, non-synaptic adaptations, and network state-dependent processes. As we further explore predictive processing, these adaptive behaviors provide an essential foundation for understanding how simplified neural networks manage information and anticipate future events.

## 4. Prediction, Deviance Detection, and the Free Energy Principle

The free energy principle and predictive coding framework propose that neural systems maintain internal models to minimize prediction errors about their sensory inputs. Under this framework, neural responses represent prediction errors - the difference between expected and actual inputs. Organisms actively minimize prediction errors through two complementary processes: updating internal models to better predict sensory inputs and selecting actions that confirm these predictions. This principle helps explain phenomena like mismatch negativity (MMN), where the brain produces enhanced responses to stimuli that violate statistical regularities, representing prediction error signals in sensory processing hierarchies.

In dissociated neuronal cultures, evidence for predictive processing comes from multiple experimental approaches. Early evidence for differential processing of frequent and rare stimuli came from Eytan et al. (2003), who showed that cortical networks could selectively adapt to different stimulation patterns using multi-electrode arrays. Their work demonstrated that neurons attenuated responses to frequent stimuli while enhancing responses to rare events. Through careful pharmacological manipulations, they showed this selective adaptation depended on both excitatory synaptic depression and GABAergic inhibition, though their findings likely primarily reflect stimulus-specific adaptation (SSA) mechanisms rather than true prediction error signaling. The distinction between SSA and genuine deviance detection became clearer through subsequent work. While SSA reflects passive reduction in responses to repeated stimuli through synaptic depression, true deviance detection requires active comparison between predicted and actual inputs. Kubota et al. (2021a) provided preliminary evidence for genuine prediction error detection using high-density CMOS arrays. By implementing both oddball paradigms and many-standards control conditions, they demonstrated that deviant responses were enhanced beyond what SSA alone would predict. These paradigms and their results are summarized in **Figure 3**, which illustrates the experimental setup and neuronal responses. **Figure 3A** shows the electrode map of the high-density CMOS microelectrode array, highlighting the spatial distribution of stimulating and recording sites. **Figure 3B** details the stimulation protocols used in the oddball and many-standards control paradigms, demonstrating how the alternation of standards and deviants elicits differential responses. **Figure 3C** compares neural responses to standard and deviant stimuli, with raster plots and population peristimulus time histograms (p-PSTHs) revealing that deviant stimuli elicit stronger and more widespread responses than standards, particularly in the late response phase. Recent work (Zhang et al. 2025, in press) has solidified these findings using additional controls and larger sample sizes to confirm that the enhanced mismatch responses are not artifacts of simpler mechanisms like stimulus-specific adaptation. These findings were particularly robust in demonstrating mismatch responses dependent on NMDA receptor function, mirroring their role in MMN generation in intact brains and highlighting the critical role of synaptic plasticity in neural prediction. Additionally, this study showed that cultured networks can detect violations of complex statistical regularities, providing further evidence for their sophisticated mismatch responses and sensitivity to sequence predictability, similar to capabilities previously observed only in intact cortex (Yaron et al., 2012). The findings suggest these basic networks possess intrinsic capabilities for statistical learning and prediction. The mechanistic basis for deviance detection has been illuminated through computational modeling. Kern and Chao (2023) demonstrated that the interaction between two forms of short-term plasticity—synaptic short-term depression (STD) and threshold adaptation (TA)—can explain how neural networks achieve deviance detection. Their work showed that threshold adaptation alone enables basic deviance detection by reducing responses to frequent stimuli while maintaining sensitivity to unexpected inputs. However, the combination of TA with synaptic short-term depression produces enhanced deviance detection through synergistic effects: local synaptic fatigue from STD amplifies the global recovery mediated by TA. This mechanism allows networks to effectively encode predictable patterns while maintaining heightened sensitivity to novel stimuli, providing a computational foundation for understanding how neural circuits implement prediction error detection.

Strong evidence for predictive processing in cultured networks comes from studies demonstrating Bayesian inference capabilities. Isomura et al. (2015) showed that cortical neurons in culture could perform blind source separation using a microelectrode array (MEA) system. By delivering mixed stimuli containing distinct patterns, they demonstrated that rat cortical neurons could develop selective responses to specific stimulus aspects through Hebbian plasticity, distinguishing individual sources within the mixed inputs. This work provided initial support for free energy minimization in simplified neural circuits. Building on this foundation, Isomura and Friston (2018) explored how neuronal cultures perform inference about hidden causes in their sensory environment. By stimulating cortical neurons with probabilistic input patterns, they observed neurons developing functional specialization - selectively responding to certain hidden sources within mixed stimuli. This selective response pattern aligned with Bayesian inference under the free energy principle, as neurons refined their responses based on accumulated evidence regarding the sources generating their inputs. Recent work Isomura et al. (2023) provided the most direct evidence yet by demonstrating that dissociated neuronal networks perform variational Bayesian inference. Using an MEA to deliver structured stimuli composed of two hidden sources, they observed that neuronal networks adapted their responses through synaptic adjustments, functioning as probabilistic beliefs about the sources. Notably, pharmacological manipulation of network excitability altered these "prior beliefs," offering direct evidence for variational free energy minimization in simplified neural systems.

The relationship between prediction and memory formation has been illuminated by (Lamberti et al., 2023), who demonstrated that focal electrical stimulation generates more effective long-term memory traces compared to global stimulation. Using detailed analysis of network responses, they showed that spatially specific activation patterns enhance the network's ability to predict future inputs. This suggests that localized stimulation allows networks to build more accurate predictive models through targeted synaptic modifications. Their follow-up study (Lamberti et al., 2024) provided mechanistic insights by revealing that NMDA receptor activity is crucial for stabilizing these memory traces and improving prediction, demonstrating how synaptic plasticity enables networks to build and refine their predictive models.

These findings demonstrate that even simplified neuronal networks can implement core aspects of predictive processing - from basic prediction error detection to sophisticated Bayesian inference. While the exact mechanisms may differ from intact brains, the evidence suggests that prediction is a fundamental feature of neural computation that can be studied effectively in reduced preparations. Understanding how these basic circuits implement prediction may inform both theories of brain function and development of artificial systems incorporating similar principles.

**5. Goal-Directed Behavior**

Dissociated neuronal cultures, when integrated with embodied systems, provide a powerful model for studying goal-directed behavior. Potter et al. (1997) pioneered this field by introducing the 'Animat in a Petri Dish' concept, combining cultured neural networks with real-time computing environments. Using multi-electrode arrays (MEAs) and advanced imaging techniques, they established a paradigm where network activity controlled a simulated animal ('animat') while receiving sensory feedback through electrical stimulation. This groundbreaking work demonstrated the potential for studying learning and memory in simplified neural networks through feedback-driven interaction with their environment.

DeMarse et al. (2001) built upon this foundation by demonstrating that cultured networks could control a simulated aircraft's pitch and roll in a virtual environment, showing that these cultures could learn to maintain flight stability over time. Potter et al. (2004) further advanced the field by introducing "Hybrots" (hybrid neural-robotic systems), where cultured networks served as "brains" for robotic systems. This approach addressed limitations of traditional in vitro systems by providing sensory inputs and motor outputs through closed-loop interaction.

A systematic investigation of these systems emerged through a series of complementary studies. Chao et al. (2005) demonstrated that random background stimulation could stabilize synaptic weights after tetanization in both simulated and living networks, preventing spontaneous bursts from disrupting learned patterns. They developed novel analytical tools, further refined in Chao et al. (2007), including the Center of Activity Trajectory (CAT) to better detect and analyze network plasticity. This work provided the methodological foundation for more complex behavioral studies.

Chao et al. (2008a, 2008b) demonstrated how simulated neural networks could be shaped for adaptive, goal-directed behavior. Using leaky integrate-and-fire neurons inspired by cortical cultures, they created a closed-loop system where an animat learned to move and remain within specific target areas. Their work revealed several key principles: random background stimulation was crucial for maintaining network stability, successful adaptation required stimuli that evoked distinct network responses, and long-term plasticity through STDP was essential for learning. Building on these insights, (Bakkum et al., 2008b) made the crucial advance of implementing these principles in living neural networks. Using multi-electrode arrays, they showed how real biological networks could be trained to perform goal-directed behavior through a structured combination of context-control probing sequences (CPS), patterned training stimulation (PTS), and random background stimulation (RBS). Their success in training cultures to guide an animat toward predefined areas demonstrated that biological neural circuits could be shaped for adaptive control in real-world applications, establishing a foundation for developing neuroprosthetics and therapeutic interventions. This work provided definitive evidence that living neuronal networks could be systematically trained to perform specific behaviors through carefully designed stimulation protocols.

Tessadori et al. (2012) further explored modular network architectures, showing that hippocampal neurons divided into distinct compartments could enhance goal-directed behavior. Their virtual robot avoided obstacles in an arena by interfacing with the neuronal culture, with tetanic stimulation applied to reinforce successful movements. Modular networks exhibited more structured and selective neural activity, improving the robot's performance compared to random networks.

Recent advances have explored new computational paradigms in these systems. Masumori et al. (2020) introduced the concept of "neural autopoiesis," showing how networks can regulate self-boundaries through stimulus avoidance behaviors. Their work revealed how networks adaptively distinguish between controllable and uncontrollable inputs, providing insights into neural self-organization and adaptation. Yada et al. (2021) demonstrated physical reservoir computing with FORCE learning in living neuronal cultures. **Figure 4** illustrates this closed-loop system, where cortical neurons cultured on a microelectrode array (MEA) generate spiking activity processed via FORCE learning to create coherent signals. **Figure 4A** shows the system's design, including optical stimulation using a digital micromirror device (DMD) for feedback. **Figure 4B** demonstrates the robot navigation task, where neuronal activity controls a robot navigating through a maze toward a goal (highlighted in yellow), with electrical stimulation applied when obstacles are encountered. Feedback from the environment guides the robot's trajectory, highlighting how intrinsic neural dynamics, coupled with real-time learning algorithms, enable adaptive task performance. This work underscores the potential of embodied neuronal networks for solving goal-directed tasks without additional external learning mechanisms.

Kagan et al. (2022) made a significant advance by demonstrating that dissociated neuronal cultures could rapidly adapt to controlling a paddle in a simplified "Pong" game. Using a high-density multi-electrode array (HD-MEA) with 26,400 electrodes, the system provided real-time feedback to neurons, which were able to adjust their firing patterns within minutes. Their latest work (Khajehnejad et al., 2024) compared the learning efficiency of biological neurons with deep reinforcement learning (RL) algorithms, revealing that neurons could learn faster in environments with limited training data, highlighting their unique adaptability.

Moving forward, these works open new opportunities for exploring more complex tasks in embodied neural systems, though questions about the intelligence or sentience of these behaviors remain (Balci et al., 2023). Further research could involve more intricate feedback systems and multi-compartment setups, to deepen our understanding of neuronal plasticity and prediction in embodied systems, with potential applications in neuroprosthetics, robotics, and bio-hybrid systems.

**6. Insights for Artificial Neural Networks and Neuromorphic Systems**

Research into dissociated neuronal cultures has become increasingly relevant for designing neuromorphic computing systems that address traditional computing limitations. The scale of this challenge is striking: Marković et al. (2020) highlight that training a single state-of-the-art natural language processing model on conventional hardware consumes energy equivalent to running a human brain for six years. In contrast, biological neural networks perform complex computations with remarkable energy efficiency, requiring approximately 20 W for the entire human brain. Beyond energy savings, neuronal cultures offer a paradigm where computation and memory coexist within the same substrate, which may interface directly with biological systems (Gentili et al., 2024). The computational properties of neuronal cultures, detailed in earlier chapters, suggest principles for artificial system design. From their self-organization toward critical states optimizing information flow (Chapter 2), to their demonstrations of adaptability and learning (Chapter 3), deviance detection, and predictive coding (Chapter 4), these networks display capabilities crucial for efficient information processing, adaptation, and prediction.

Early studies revealed fundamental aspects of temporal processing in neural systems. Buonomano and Maass (2009) demonstrated how cortical networks process spatiotemporal information by encoding temporal sequences through transient activity patterns. They showed how recurrent connections and short-term synaptic plasticity enable sequence recognition and prediction. Nikolić et al. (2009) revealed that neurons in the visual cortex retain fading memories of stimuli for several hundred milliseconds, using multi-electrode recordings to show that this supports sequential processing. Later work by Enel et al. (2016) extended this by demonstrating reservoir computing properties in the prefrontal cortex, showing how high-dimensional dynamics allow adaptive decision-making through mixed selectivity, while Seoane (2019) examined reservoir computing from an evolutionary perspective. Various approaches have emerged for implementing neural computation in artificial systems. Abbott et al. (2016) tackled challenges in building functional spiking networks, emphasizing stable excitation-inhibition balance and scalable training mechanisms. Learning strategies in artificial systems have also drawn from these findings: Diehl and Cook (2015) demonstrated unsupervised learning in spiking networks with STDP, using spike rates to classify MNIST digits with competitive accuracy, while Nicola and Clopath (2017) introduced FORCE training, stabilizing chaotic network dynamics to reproduce complex temporal sequences like oscillations and trajectories. Reservoir computing applications in neuronal cultures have revealed increasing sophistication in computational capabilities. Dockendorf et al. (2009) demonstrated that cultured networks could act as liquid state machines, effectively separating input patterns with high-frequency stimulation. Kubota et al. (2019) identified the echo state property in cultured networks, which is crucial for maintaining short-term memory and processing temporal information. They demonstrated that cultured networks could hold transient states that preserve past inputs while enabling flexible processing of new information. Using high-density multielectrode arrays, they systematically tested various inter-pulse intervals (IPIs) and found that the optimal range, particularly between 20 and 30 ms, maximized reproducibility and differentiation of neural responses. This optimal timing reflects the networks' ability to encode and process information with minimal interference or loss. (Kubota et al., 2021) expanded on this work by quantifying the networks' information processing capacity (IPC), a comprehensive metric capturing their computational versatility. Their analysis revealed how memory capacity and generalization ability in these networks depend on specific stimulation patterns, bridging physical reservoir computing with living neural systems. (Suwa et al., 2022) demonstrated that dissociated cortical cultures possess both first-order IPC (linear memory of past inputs) and second-order IPC (interactions of past inputs), enabling them to perform arithmetic and logical operations on previous stimuli. Using high-density CMOS microelectrode arrays, they quantified the computational capabilities of these cultures, showing how spatiotemporal neural activity supports advanced processing tasks. Their findings underscore the capacity of neuronal cultures to serve as living reservoirs for complex computations, bridging the gap between biological and artificial systems while providing insights into the intrinsic mechanisms underlying neural computation. Ikeda et al. (2023) further refined these insights by investigating the dynamic interaction between evoked and spontaneous activities. They highlighted the importance of evoked response intensity and identified conditions, such as a 30-ms IPI, that optimize IPC. These findings collectively underscore the potential of cultured networks to act as robust and adaptable computational substrates, providing critical benchmarks for designing bio-inspired computing architectures. (Subramoney et al., 2019, 2021) proposed the "Learning-to-Learn" framework, enabling spiking neural networks to adapt rapidly to new tasks by leveraging meta-learning strategies. This framework fine-tunes internal structures using task-specific feedback, optimizing performance across diverse computational challenges. It exemplifies a scalable and efficient approach for building neuromorphic systems that can generalize and adapt to novel inputs. Ishikawa et al. (2024) integrated predictive coding principles with reservoir computing in spiking neural networks, advancing the capacity for dynamic temporal processing. Predictive coding allows networks to anticipate future inputs by leveraging past patterns, optimizing processing efficiency and reducing computational redundancy. Their work demonstrates how combining these principles enhances spiking neural networks' ability to process sensory information in real-time, particularly in tasks requiring temporal predictions and interaction with dynamic environments.

The presence of noise in biological neural systems represents not a limitation but a crucial computational resource that enables energy-efficient processing. Unlike digital computers, which require high signal margins (e.g., 0 vs 5V) to maintain adequate signal-to-noise ratios, biological networks harness noise for computation. Early studies revealed fundamental principles: Matsumoto and Tsuda (1983) showed that noise stabilizes chaotic systems by reshaping trajectories into periodic orbits, while Kirkpatrick et al. (1983) showed how noise-based optimization through simulated annealing could solve complex problems. Gassmann (1997) demonstrated noise-induced transitions between chaos and order, and Gammaitoni et al. (1998) showed how stochastic resonance could enhance weak signal detection. Anderson et al. (2000) revealed noise's role in maintaining visual contrast invariance, demonstrating how seemingly random fluctuations support robust sensory processing. A comprehensive review by Faisal et al. (2008) documented noise's pervasive and often beneficial role throughout nervous systems. Subsequent work demonstrated specific computational advantages: Habenschuss et al. (2013) showed how cortical circuits harness noise for stochastic computation, and Maass (2014) established noise as a resource for learning in spiking networks. This noise-harnessing computation represents an evolutionary adaptation that allows the brain to operate in harsh biochemical environments while maintaining energy efficiency. Recent studies have revealed specific mechanisms by which noise shapes neural computation in biological and artificial systems. (Ikeda et al., 2023) demonstrated that noise interacts with spike-timing-dependent plasticity to drive self-organized criticality in spiking neural networks. They showed that moderate noise levels optimize critical dynamics while maintaining stable synaptic structures, whereas excessive noise disrupts network stability and computational capacity. Ikeda et al. (2024) further revealed how noise-driven spontaneous activity serves broader computational functions, maintaining criticality and supporting memory consolidation through homeostatic processes. These findings suggest that incorporating controlled noise in neuromorphic systems might improve their adaptability and computational efficiency, paralleling the sophisticated use of noise observed in biological neural networks.

These developments continue to drive advances in neuromorphic computing. The computational properties observed in neuronal cultures—from criticality to noise-harnessing computation—provide blueprints for energy-efficient, adaptive architectures. The field moves toward systems that may operate synergistically with living neural tissue, combining the advantages of biological and artificial computation.

## 7. Conclusions and Future Directions

Dissociated neuronal cultures serve as powerful, simplified model systems for examining fundamental neural processes. As detailed in this review, these cultures exhibit complex dynamics characteristic of self-organized criticality and adaptive computation (Chapter 2), demonstrate learning and memory formation through various plasticity mechanisms (Chapter 3), and show predictive processing and deviance detection capabilities consistent with theoretical frameworks like the free energy principle (Chapter 4). They have also demonstrated the capacity for goal-directed behaviors in controlled, closed-loop environments, further illustrating their potential as computational models (Chapter 5). These discoveries not only enhance our understanding of biological neuronal function but also provide insights that could influence the design of future artificial neural networks and computational architectures due to the unique blend of simplicity, adaptability, and controllability found in these in vitro systems (Chapter 6).

While our current understanding of dissociated neuronal cultures is robust, several avenues remain open for deepening our knowledge and refining the practical applications of these systems. Continued advancements in microelectrode array (MEA) technology are expected to enable more precise recordings and manipulations of neuronal activity in dissociated cultures, allowing for an even deeper exploration of their computational properties. Ravula et al. (2007) pioneered early work in this direction by developing a microfabricated compartmentalized culture system that leveraged microfluidics for precise spatial and temporal control over neuronal microenvironments. This design improved upon traditional compartment-based methods by facilitating reliable fluid isolation and collagen-guided axonal growth, thus enabling simultaneous electrophysiological recordings and drug exposures. Improvements in the spatial and temporal resolution of MEAs may further clarify how specific patterns of connectivity and synaptic plasticity underlie adaptive computations and dynamic behavior in neuronal networks.

New MEA designs featuring modularity—such as multi-well MEAs that physically and functionally separate different neuronal populations—also hold the potential to recreate hierarchical and modular network structures in vitro. Bisio et al. (2014), for instance, demonstrated how modular networks grown on polydimethylsiloxane (PDMS) structures can exhibit higher firing rates during early development and display unique synchronization properties compared to uniform networks, shedding light on hierarchical organization. Building on this foundation, Joo and Nam (2019) introduced an agarose-based microwell patterning method, enabling the recording of slow-wave activity from micro-sized neural clusters while preserving high-frequency spiking information. Negri et al. (2020) refined protocols for multi-well MEA experiments—providing a spike-sorting pipeline and statistical methodologies to improve reproducibility—while also highlighting the importance of proper experimental design. More recent work by (Gladkov et al., 2017, 2021) and Duru et al. (2022) has further extended the engineering of biological neural networks by integrating microstructures with high-density CMOS arrays. These approaches not only confine axonal outgrowth to specific channels, creating reproducible unidirectional connectivity, but also offer subcellular-resolution recordings of directed spike propagation. Finally, (Sumi et al., 2023) revealed how increasing network modularity enhances reservoir computing performance in biological neuronal networks, enabling improved classification accuracy in both spatial and temporal tasks. Their findings suggest that structured designs can foster dynamic states conducive to advanced processing and short-term memory.

Looking ahead, research into three-dimensional neuronal culture systems and brain organoids may reveal how increasing the complexity of these in vitro models affects network organization and computation. By introducing additional layers of structural and functional complexity, researchers can investigate how hierarchical connectivity and layered processing influence predictive coding, learning, and memory. Such 3D cultures and organoids more closely mimic the architecture of in vivo brain tissue, potentially providing deeper insights into complex cognitive functions and developmental processes. However, even as complexity increases, the fundamental simplicity and controllability of these cultures remain advantageous, allowing precise manipulation and observation of the network's activity (Hogberg et al., 2013; Lancaster et al., 2013; Clevers, 2016; Smirnova and Hartung, 2022, 2024).

Future studies will likely explore how the principles uncovered in dissociated neuronal cultures generalize to more complex neural systems. While introducing 3D structures and organoids adds realism, it is the balance between complexity and controllability that makes these models so valuable. Researchers will need to maintain the simplicity that allows for precise control and manipulation, ensuring that the systems remain tractable for in-depth investigations of network function. By carefully scaling complexity, it is possible to examine how additional layers of organization and connectivity influence predictive processing and adaptive computation without losing the crucial benefits of simplicity. There is also significant potential for an increased synergy between experimental neuroscience and computational modeling. As our ability to record and manipulate neuronal activity improves, so does our capacity to develop and refine computational models that can predict network behavior. These models can, in turn, guide experimental interventions, allowing researchers to probe network function more systematically. This iterative process between experimentation and modeling may help identify the principles underpinning self-organization, learning, and prediction in neural networks and aid in translating these insights into artificial systems.

In summary, dissociated neuronal cultures remain an invaluable model system for exploring fundamental aspects of neuronal function and computation, particularly the mechanisms underlying self-organized prediction. They have proven essential in examining how networks self-organize, learn, and adapt, providing a simplified and controllable environment to study complex neural phenomena that underlie predictive processing. As researchers continue to balance the simplicity of these systems with increasing complexity—our understanding will deepen further. These insights not only elucidate how biological brains function through prediction and adaptation but also inspire the next generation of computational architectures and neurotechnological applications.

**Figure legends**

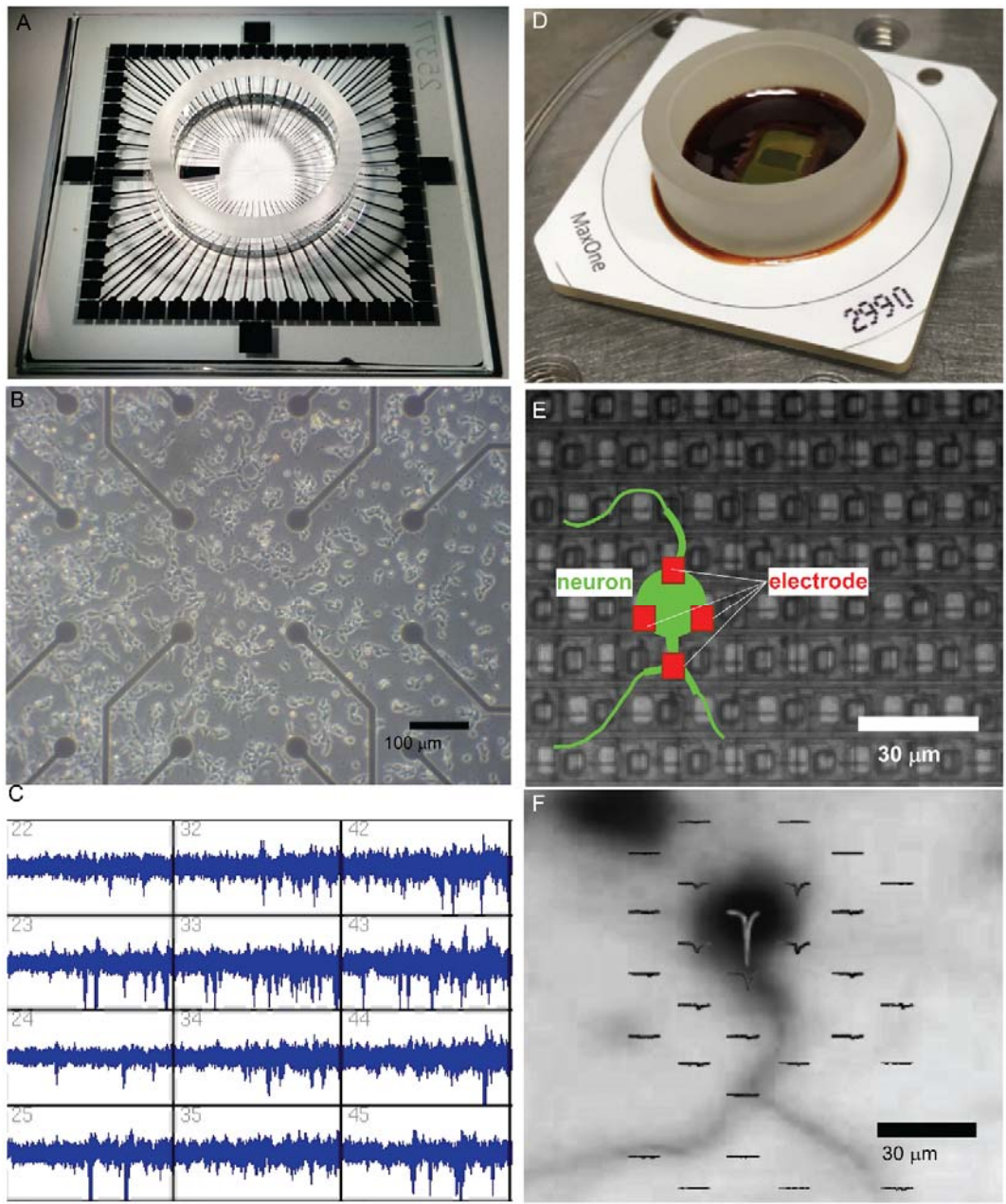

**Figure 1: Evolution of microelectrode array (MEA) technology for studying neuronal networks.** (A) The MEA system, featuring a transparent glass substrate with 60 microelectrodes spaced at 200 μm. This design provides sufficient spatial resolution for capturing network-level neuronal activity and allows for optical imaging of the culture. The system is capable of both extracellular recording and stimulation for long-term culture studies. (B) Bright-field microscopy of dissociated neuronal cultures grown on the MEA platform. The electrode array beneath the neuronal layer supports the self-organization of functional networks while enabling the simultaneous observation of culture morphology and recording of extracellular signals. Scale bar

= 100 µm. (C) Extracellular spike recordings from MEA, demonstrating its capacity to capture neuronal activity from multiple electrodes simultaneously. The recording resolution and electrode layout enable the analysis of network activity patterns and dynamic behaviors. (D) High-dense CMOS-based MEA system (MaxOne), incorporating 26,400 platinum electrodes with a 17.5 µm pitch. This CMOS-MEA provides subcellular spatial resolution for recording and stimulation, enabling the detailed investigation of localized neuronal activity and network interactions. (E) Schematic overlay of a neuron (green) interacting with electrodes (red) on a CMOS- MEA. The figure illustrates how neuronal somas and processes align with the electrode array. The red electrodes in close proximity to the soma demonstrate the ability of high-density CMOS arrays to monitor and stimulate activity at a single-cell resolution. The scale bar indicates the high spatial resolution provided by this system, with electrodes spaced at approximately 17.5 µm. (F) CMOS-MEA monitoring an action potential generated from the soma. Immunostaining image of a neuron on the CMOS MEA is overlaid with spatially localized extracellular spike sources. The high-density electrode array enables the resolution of neuronal activity at subcellular precision, revealing fine-scale functional properties of single neurons and their interactions with the network. Scale bar = 30 µm.

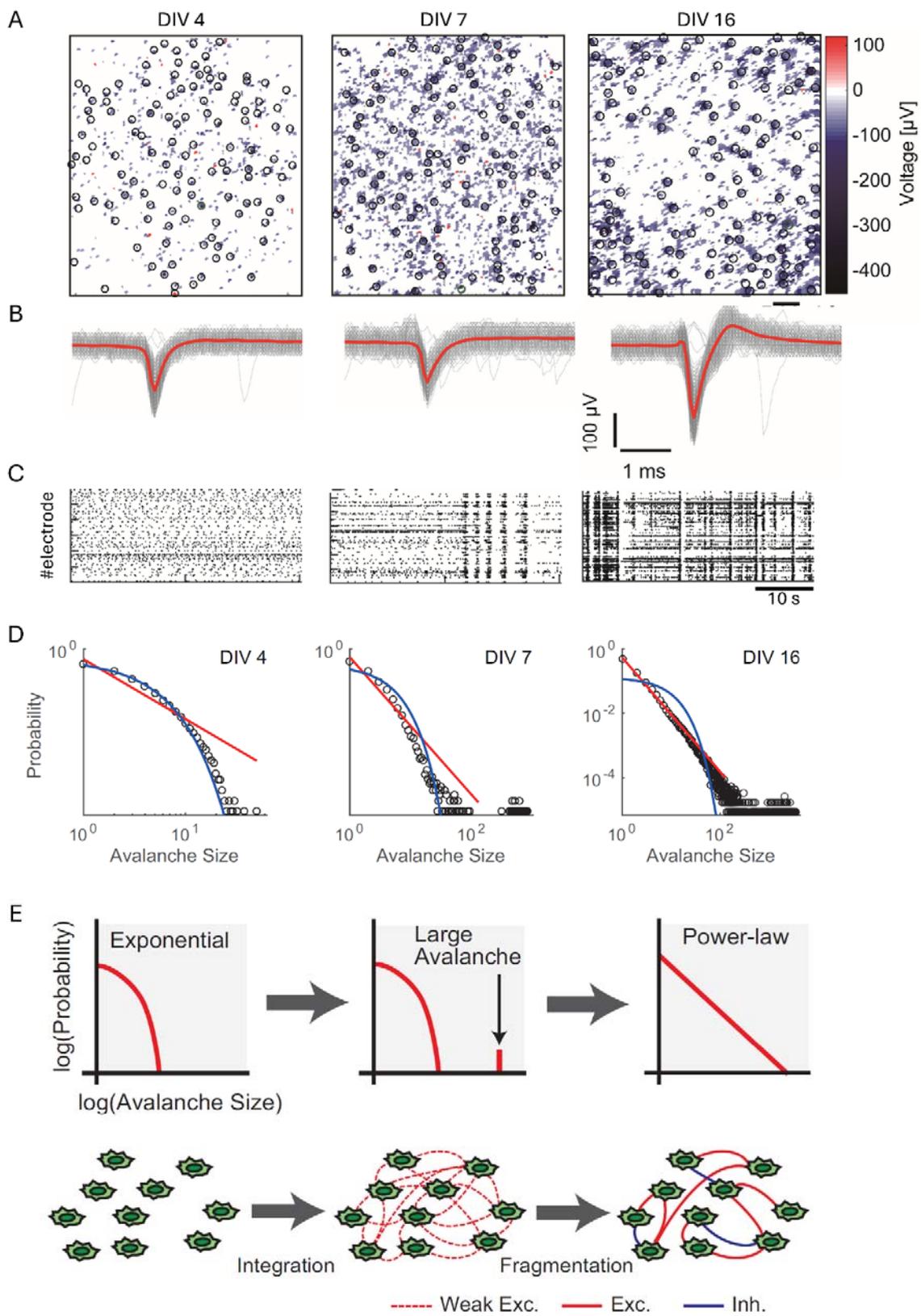

**Figure 2**: **Developmental transition toward self-organized criticality (SoC) in dissociated neuronal cultures. (A)** Spatial maps of action potential amplitudes recorded using high-density CMOS MEAs at different developmental stages: 4 days in vitro (DIV), 7 DIV, and 16 DIV. Black circles mark recording sites, and the heatmap represents voltage amplitudes (color scale: −400 to 100 µV). Scale bar = 200 µm. **(B)** Representative spike waveforms recorded at selected electrodes (indicated by black circles in (A)) across developmental stages. Grey lines depict raw spike traces, while red lines indicate averaged spike waveforms. Scale bars = 1 ms, 100 µV. **(C)** Raster plots of spontaneous spiking activity from 120 s of recorded data for the same cultures at 4, 7, and 16 DIV, illustrating the emergence of synchronized bursts over time. **(D)** Log-log plots of neuronal avalanche size distributions at 4, 7, and 16 DIV. Exponential distributions dominate early development (4 DIV), while bimodal distributions emerge at 7 DIV, and power-law distributions characteristic of SoC appear by 16 DIV. Fitted red lines represent power-law distributions, and blue lines indicate exponential fits. **(E)** Schematic representation of the integration-fragmentation model for SoC emergence. Initially, neurons form weak excitatory connections, generating exponential distributions. Large-scale avalanches emerge as connectivity strengthens, leading to a bimodal distribution. Finally, synaptic pruning and the balance of excitation and inhibition result in diverse avalanche sizes distributed according to a power-law. Figure reproduced from Yada et al. (2017), "Development of neural population activity toward self-organized criticality," Neuroscience, 343, 55–65.

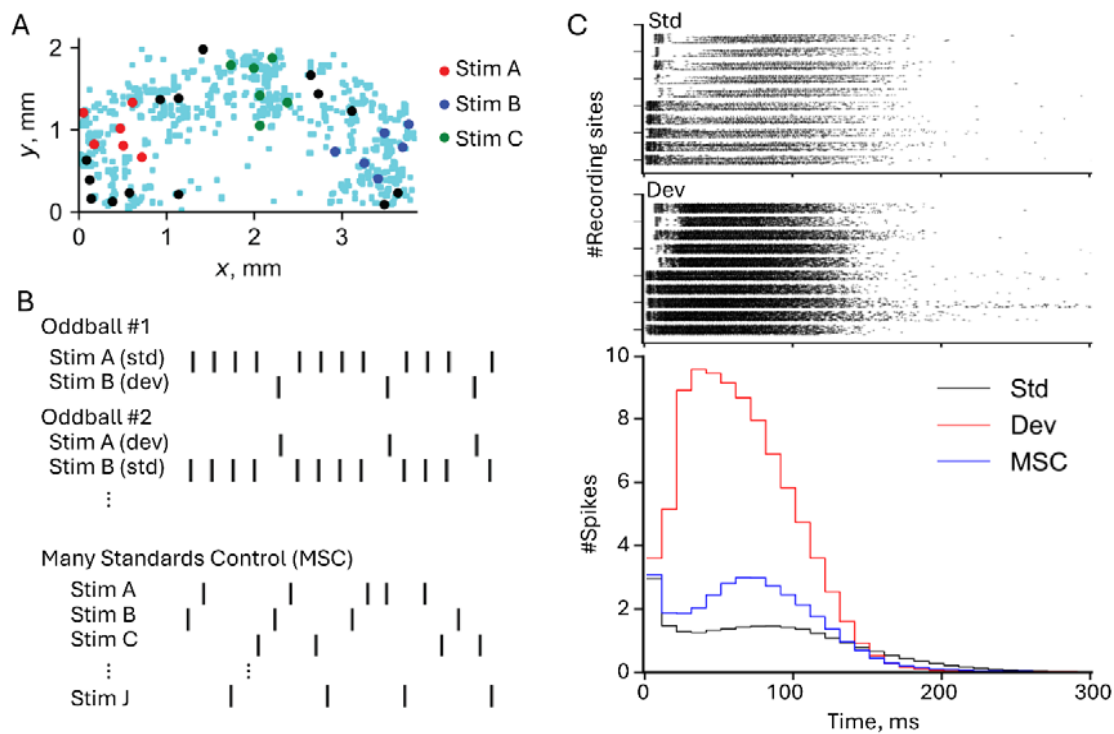

**Figure 3: Experimental paradigm and deviance detection responses in neuronal networks.** (A) Electrode map from a high-density CMOS microelectrode array showing the spatial distribution of stimulating electrodes (red, blue, and green dots for Stim A, Stim B, and Stim C, respectively) and recording sites (light blue dots). Stimuli were delivered at specific locations to investigate network responses. (B) Stimulation protocols used in the oddball and many standards control (MSC) paradigms. In the oddball paradigm, Stim A and Stim B were alternated as standard (std) and deviant (dev) stimuli. In the MSC paradigm, multiple stimuli (Stim A, Stim B, Stim C, etc.) were presented in random order to eliminate expectations of repetition. (C) Top: Raster plots showing neural responses to standard (top) and deviant (bottom) stimuli. Each row corresponds to a recording site, and black dots indicate spike times relative to the stimulus onset. Deviant stimuli elicited stronger and more widespread responses compared to standards. Bottom: Population peristimulus time histograms (p-PSTHs) comparing the number of spikes per time bin across conditions. Deviant stimuli (red line) evoke higher firing rates and longer-lasting responses than standard (black line) and MSC (blue line) conditions, particularly in the late response phase (30-100 ms). Figure modified from Kubota et al. (2020).

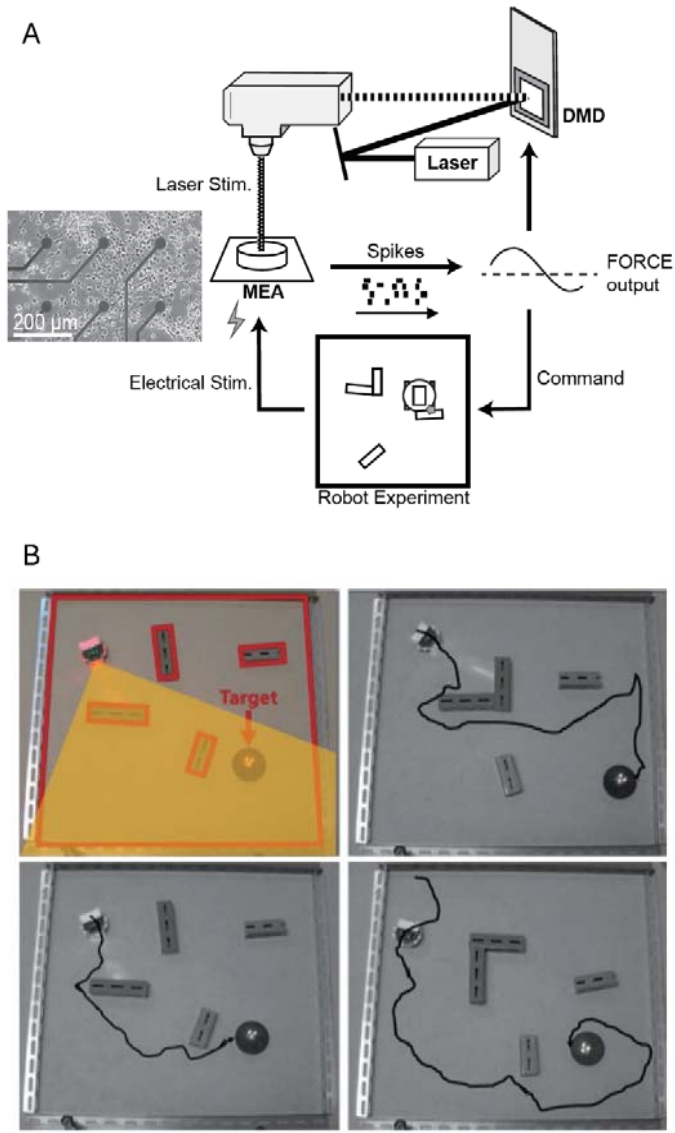

**Figure 4. Closed-loop system for goal directed behavior using a living neuronal culture.** (A) Schematic representation of the closed-loop system. Cortical neurons cultured on a microelectrode array (MEA) generate spiking activity, which is recorded and processed via FORCE learning to create a coherent signal. For FORCE learning, the feedback to the neuronal network is provided via optical stimulation (using a digital micromirror device, DMD). (B) Robot navigation task. Representative trajectories of a robot in a maze with obstacles toward a designated goal (target zone highlighted in yellow) are shown. The robot's movements are controlled by neuronal activity, with FORCE learning enabling adaptive task performance. Electrical stimulation is applied when the robot hit an obstacle. Feedback from the

environment—through optical and electrical stimulation—guides the robot's trajectory toward the goal. Figure adapted from Yada et al. (2021), "Physical reservoir computing with FORCE learning in a living neuronal culture," Applied Physics Letters, 119, 173701.

**Ethics statement**

This study is a review of previously published research and does not involve any experiments with human participants or animals performed by the authors. All referenced studies have adhered to their respective ethical standards as stated in their publications.

**Author Contributions**

AY drafted the manuscript and organized its overall structure. ZZ conducted literature review and provided critical feedback. DA contributed feedback and refinements across multiple sections of the manuscript. TIS contributed feedback and refinements on deviant detection. ZC provided substantial input throughout the manuscript, with significant contributions to the section on goal-directed behavior. HT supervised the project, provided key references, contributed to the theoretical framework, prepared the figures, and ensured the manuscript's coherence and finalization. All authors contributed to the work and approved the final version of the manuscript.

**Funding**

This work is partly supported by JSPS KAKENHI (23H03465, 23H04336, 24H01544, 24K20854), AMED (24wm0625401h0001), JST (JPMJPR22S8), the Asahi Glass Foundation, and the Secom Science and Technology Foundation.

**Acknowledgments**

The authors acknowledge the use of ChatGPT (OpenAI, version 4o) for assistance in language editing.

**Conflict of Interest**

*The authors declare that the research was conducted in the absence of any commercial or financial relationships that could be construed as a potential conflict of interest.*